# Cosmology: small scale issues revisited

Joel R. Primack[a]

[a]Physics Department, University of California, Santa Cruz, CA 95064 USA

**Abstract.** The abundance of dark matter satellites and subhalos, the existence of density cusps at the centers of dark matter halos, and problems producing realistic disk galaxies in simulations are issues that have raised concerns about the viability of the standard cold dark matter (ΛCDM) scenario for galaxy formation. This article reviews these issues, and considers the implications for cold vs. various varieties of warm dark matter (WDM). The current evidence appears to be consistent with standard ΛCDM, although improving data may point toward a rather tepid version of ΛWDM – tepid since the dark matter cannot be very warm without violating observational constraints. (This is a substantially updated and expanded version of my talk at the DM08 meeting at Marina Del Rey, arXiv:0902.2506.)

## DARK MATTER IS OUR FRIEND

Dark matter preserved the primordial fluctuations in cosmological density on galaxy scales that were wiped out in baryonic matter by momentum transport (viscosity) as radiation decoupled from baryons in the first few hundred thousand years after the big bang. The growth of dark matter halos started early enough to result in the formation of galaxies that we see even at high redshifts $z > 6$. Dark matter halos provide the gravitational potentials within which stable structures formed in the universe. In more recent epochs, dark matter halos preserve these galaxies, groups, and clusters as the dark energy tears apart unbound structures and expands the space between bound structures such as the Local Group of galaxies. Thus we owe our existence and future to dark matter.

Cold dark matter theory [1] including cosmic inflation has become the basis for the standard modern ΛCDM cosmology, which is favored by analysis of the available cosmic microwave background data and large scale structure data over even more complicated variant theories having additional parameters [2]. Most of the cosmological density is nonbaryonic dark matter (about 23%) and dark energy (about 72%), with baryonic matter making up only about 4.6% and the visible baryons only about 0.5% of the cosmic density. The fact that the universe is mostly invisible, with the dominant contributions to the cosmic density being dark energy and dark matter, suggests a popular name for the modern standard cosmology: the "double dark" theory, as Nancy Abrams and I proposed in our recent book about modern cosmology and its broader implications [3].

Despite a long history [4] of observation and theory, the physical nature of dark matter remains to be discovered. The two most popular ideas concerning the identity of the dark matter particles are the lightest supersymmetric partner particle [5], also called supersymmetric weakly interacting massive particle (WIMP) [6], and the cosmological axion, recently reviewed in [7]. WIMPs and axions are the two dark matter candidate particles that are best motivated, in the sense that they are favored by other considerations of elementary particle theory.

Supersymmetry remains the best idea for going beyond the standard model of particle physics. It allows control of vacuum energy and of otherwise unrenormalizable gravitational interactions, and thus may allow gravity to be combined with the electroweak and strong interactions in superstring theory. Supersymmetry also allows for grand unification of the electroweak and strong interactions, and naturally explains how the electroweak scale could be so much smaller than the grand unification or Planck scales (thus solving the "gauge hierarchy problem"). The connection of supersymmetry breaking with electroweak symmetry breaking leads to the expectation that the supersymmetric WIMP mass will be in the range of about 100 to 1000 GeV. This also leads to the "WIMP miracle," the fact that the WIMP cosmological density has approximately the observed value.

Axions remain the best solution to the CP problem of the standard SU(3) gauge theory of strong interactions, although it is possible that the axion exists and solves the strong CP problem but makes only a negligible contribution to the dark matter density.

Many other particles have been proposed as possible dark matter candidates, even within the context of supersymmetry. An exciting prospect in the next few years is that experimental and astronomical data may point toward specific properties of the dark matter particles, and may even enable us to discover their identity. The present paper is concerned with potential problems for CDM and clues to the nature of the dark matter from astronomical data such as substructure within dark matter halos, especially subhalos and satellites, central cusps, and angular momentum issues.

## SUBHALOS AND SATELLITES

It at first seemed plausible that the observed bright satellite galaxies are hosted by the most massive subhalos of the dark matter halo of the central galaxy, but this turned out to predict too large a radial distribution for the satellite galaxies. Andrey Kravtsov and collaborators [8] proposed instead that bright satellite galaxies are hosted by the subhalos that were the most massive when they were accreted. This hypothesis appears to predict much better the observed radial distribution of galaxies within clusters, which roughly follow the dark matter distribution. It also explains naturally, based on tidal heating, why nearby satellites are dwarf spheroidals (dSph) while more distant ones are a mix of dSph and dwarf irregular (dIrr) galaxies [8]. Such ideas can also explain the compact radial distribution of the Local Group satellites [9], but they do not account readily for the fact that the Milky Way satellites lie mostly in a plane perpendicular to the galactic disk – which may just be a local peculiarity.

An issue that is still regularly mentioned by observational astronomers (e.g. [10]) as a problem for ΛCDM and a possible argument in favor of ΛWDM is the fact that many fewer satellite galaxies have been detected in the Local Group than the number of subhalos predicted. But improving theory and the recent discovery of many additional satellite galaxies around the Milky Way and the Andromeda galaxy suggest that this may turn out not to be a problem for ΛCDM after all, as discussed in detail in a recent review by Kravtsov [9]. As Figure 1(a) shows, it is only below a circular velocity ~30 km s$^{-1}$ that the number of dark matter halos definitely begins to exceed the number of observed satellites. Figure 1(b) shows that suppression of star formation in small dwarf galaxies after reionization can account for the observed satellite abundance [11] in ΛCDM, as suggested by several authors [12-15], although the extended star formation histories [16,17] show that star formation continued in these galaxies long after reionization. It remains to be seen whether better understanding of baryonic physics can explain the recent discovery [18] that all the local faint satellites have roughly the same dynamical mass $m_{0.3}$ within their central 0.3 kpc of about $10^7$ M$_\odot$ despite having a large range of ~$10^4$ in luminosity. The inner parts of dark matter halos are formed rather early and reflect the density of the universe then; this implies that halos with a wide range of masses will all have about the required density [9]. Including tidal stripping may strengthen this argument, since more massive halos are less concentrated and thus more affected by tides [19]. Semi-analytic models and hydrodynamical simulations that suppress star formation increasingly in lower-mass halos do seem able to reproduce this and other observed features of the satellites [20-23]. Properties such as metallicity of the newly discovered ultra-faint dwarf satellite galaxies appear to continue [24] the scaling relations discovered earlier [25], with metallicity decreasing with luminosity. This supports the interpretation of these objects as dwarf galaxies with very high mass-to-light ratios, but explaining such observations in detail is a challenge [26-28] for theories of the formation of satellite galaxies. As deeper observations probe for faint dwarf galaxies at larger radii from the Milky Way, ΛCDM predicts that many more, perhaps hundreds, will be discovered [29]. Some of those at larger distances may reside in dark matter halos that have suffered less tidal stripping.

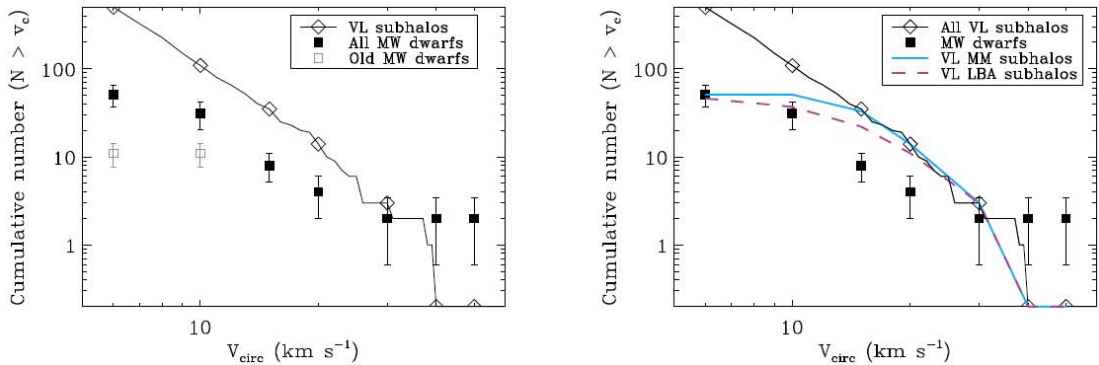

**FIGURE 1.** (a) Cumulative number of Milky Way satellite galaxies as a function of halo circular velocity, assuming Poisson errors on the number count of satellites in each bin. The filled black squares include the new circular velocity estimates from [11], who follow [30] and use V$_{circ}$ = √3 σ. Diamonds represent all subhalos within the virial radius in the first Via Lactea simulation [31]. (b) Effect of reionization on the missing satellite problem. The lower solid curve shows the circular velocity distribution for the 51 most massive Via Lactea subhalos if reionization occurred at $z$ = 13.6, the dashed curve at $z$ = 11.9, and the dotted curve at $z$ = 9.6. (Figures from [11].)

Although the abundance of nearby small satellite galaxies may be consistent with ΛCDM as we have just discussed, there may be a different problem accounting for the abundance of faint galaxies in voids. Peebles [32] had claimed that this would be a serious problem for ΛCDM, but Tinker and Conroy [33] recently compared the same data to a halo occupation distribution model of galaxy formation and showed that there was no discrepancy. However, Tikhonov and Klypin [34] recently analyzed a survey that went much fainter, and found that ΛCDM appears to overpredict the number of faint galaxies by about an order of magnitude. It remains to be seen if this discrepancy can be explained by some physical process leading to inefficient star formation in void halos – for example, because they collapsed after the reionization epoch.

Observations and Jeans analysis of the bright Milky Way dSph satellites give density profiles that are better fit by a NFW dark matter halo than a halo with a constant density core, and imply fairly strong lower limits on their central mass densities of ~1 $M_\odot$ pc$^{-3}$ [35]. Hogan and Dalcanton [36] introduced the parameter $Q = \rho/\sigma^3$ as an estimate of the coarse-grained phase-space density of the dark matter in galaxy halos. Liouville's theorem implies that observed values of Q set a hard lower limit on the original phase-space density of the dark matter. All of the galaxies except UMa I, CVn I, and Hercules have $Q > 10^{-3}$ $M_\odot$ pc$^{-3}$ (km s$^{-1}$)$^{-3}$, about an order of magnitude improvement compared to the previously known bright dSphs. The subhalos in the Via Lactea II simulation [37] that could host Milky Way satellites have densities and phase space densities comparable to these values. The relatively high Q lower limit places significant limits on non-CDM dark matter models; for example, it implies that the mass $m_{WDM}$ of a thermal WDM particle must be $m_{WDM} > 1.2$ keV. For comparison, the HIRES Lyman-α forest data implies a 2σ thermal WDM lower limit $m_{WDM} > 1.2$ keV ($m_{WDM} > 4$ keV using the much more abundant but lower resolution SDSS Lyman-α forest data) [38].

Sterile neutrinos that mix with active neutrinos are produced in the early universe and could be the dark matter [39]. Such neutrinos would decay into X-rays plus light neutrinos, so non-observation of X-rays from various sources gives upper limits on the mass of such sterile neutrinos $m_s < 3.5$ keV. Since this upper limit is inconsistent with the 2σ lower limit $m_s > 5.6$ keV from HIRES Lyman-α forest data ($m_s > 28$ keV using the SDSS Lyman-α forest data) [38], that rules out such sterile neutrinos as the dark matter, although other varieties of sterile neutrinos are still allowed and might explain neutron star kicks [40,41].

The Via Lactea II [31], GHALO [42], and Aquarius simulations [43-45] are the highest resolution ΛCDM simulations of a Milky Way mass halo yet published, and they are able to resolve substructure even at the distance of the sun from the center of the Milky Way. An important question is whether the fraction of mass in the subhalos of mass ~$10^6 – 10^8$ $M_\odot$ is the amount needed to explain the flux anomalies observed in "radio quads" – radio images of quasars that are quadruply gravitationally lensed by foreground elliptical galaxies. A recent paper [46] based on the Aquarius simulations finds that there is probably insufficient substructure unless baryonic effects improve subhalo survivability (see next section), and I understand that the Via Lactea group is reaching similar conclusions [47]. Free streaming of WDM particles can considerably dampen the matter power spectrum in this mass range, so a WDM model with an insufficiently massive particle (e.g., a standard sterile neutrino $m_\nu < 10$ keV) fails to reproduce the observed flux anomalies [48]. In order to see whether this is indeed a serious constraint for WDM and a triumph for CDM, we need more than the small number of radio quads now known – a challenge for radio astronomers!  Radio flux anomalies can be explained by ~$10^6 - 10^8$ $M_\odot$ halos, since the radio emitting region of quasars is large.  Optical flux anomalies are probably mostly caused by stellar microlensing, since the size of the quasar optical emission region is very small, but strong infrared lenses can also be useful in constraining the dark matter substructure on scales of ~$10^6 - 10^8$ $M_\odot$ since the infrared emission region is expected to be larger.  We also need better observations and modeling of these systems to see whether subhalos are indeed needed to account for the flux anomalies in all cases [49-51].  Observing time delays between the images can help resolve such issues [52].

An additional constraint on WDM comes from reionization.  While the first stars can reionize the universe starting at redshift $z > 20$ in standard ΛCDM [53], the absence of low mass halos in ΛWDM delays reionization [54].  Reionization is delayed significantly in ΛWDM even with WDM mass $m_{WDM}$ = 15 keV [55].  The actual constraint on $m_{WDM}$ from the cosmic microwave background polarization will soon be better determined by *Planck* observations.  If the WDM is produced by decay of a higher-mass particle (e.g., the "superWIMP" scenario [56] and related ideas, reviewed in [57]), the velocity distribution and phase space constraints can be different [58,59].  MeV dark matter, motivated by observation of 511 keV emission from the galactic bulge, also can suppress formation of structure with masses up to about $10^7$ $M_\odot$ since such particles are expected to remain in equilibrium with the cosmic neutrino background until relatively late times [60].

Note finally that various authors [61-63] have claimed that ΛWDM substructure develops in simulations on scales below the free-streaming cutoff. If true, this could alleviate the conflict between the many small subhalos needed to give the observed number of Local Group satellite galaxies, taking into account reionization and feedback, and needed to explain gravitational lensing radio flux anomalies.  However Wang and White [64] recently showed that such substructure arises from discreteness in the initial particle distribution, and is therefore spurious.  New high-resolution ΛWDM simulations have been done to try to avoid these problems and include the thermal velocities [65].

As a result of the new constraints just mentioned, it follows that the hottest varieties of warm dark matter are now ruled out, so if the dark matter is not cold (i.e., with cosmologically negligible constraints from free-streaming, as discussed in the original papers that introduced the hot-warm-cold dark matter terminology [1,66]) then it must at most be rather tepid.

## CUSPS IN GALAXY CENTERS

Dark matter cusps were first recognized as a potential problem for CDM by Flores and me [67] and by Moore [68].  However, beam smearing in radio observations of neutral hydrogen in galaxy centers was significantly underestimated [69,70] in the early observational papers; taking this into account, the observations imply an inner density $\rho(r) \propto r^{-\alpha}$ with slope satisfying $0 \leq \alpha < 1.5$, and thus consistent with the ΛCDM Navarro-Frenk-White [71] slope $\alpha$ approaching 1 from above at small radius

$r$. The NFW formula $\rho_{NFW}(r) = 4\rho_s x^{-1}(x+1)^{-2}$ (where $x = r/r_s$, and the scale radius $r_s$ and the density $\rho_s$ at this radius are NFW parameters) is a rough fit to the dark matter radial density profile of pure dark matter CDM halos. The latest very high resolution simulations of pure dark matter Milky-Way-mass halos give results consistent with a power law central density with α slightly greater than 1 [37] but perhaps with indications of α decreasing at smaller radii [44]. Low surface brightness galaxies are mainly dark matter, so complications of baryonic physics are minimized but could still be important [72,73]. A careful study of the kinematics of five nearby low-mass spiral galaxies found that four of them had significant non-circular motions in their central regions; the only one that did not was consistent with α ≈ 1 [74] as predicted by ΛCDM for pure dark matter halos. The central non-circular motions observed in this galaxy sample and others could be caused by nonspherical halos [75,76]. Dark matter halos are increasingly aspherical at smaller radii, at higher redshift, and at larger masses [77-80]. This halo asphericity can perhaps account for the observed kinematics [81-84], although analysis of a larger set of galaxies suggests that this implausibly requires nonrandom viewing angles [85]. Recent observations of nearby galaxies combining THINGS HI kinematic data and Spitzer SINGS 3.6 μm data to construct mass models [86] indicate that a core-dominated halo with pseudo-isothermal central profile $\rho(r) \propto (r_0^2 + r^2)^{-1}$ is preferred over a cuspy NFW-type halo for many low-mass disk galaxies, even after correcting for noncircular motions [87]. These and other observations [88] appear to favor a kpc-size core of roughly constant density dark matter at the centers of some low-mass disk galaxies. But additional observations of small spiral galaxies and faint satellite galaxies are in progress that could affect these conclusions by clarifying the effects of systematics.

Self-consistent ΛCDM simulations of galaxies including all relevant baryonic physics, which can modify the central dark matter density distributions and thus the kinematics, will also be required to tell whether ΛCDM galaxies are consistent with these observations. Attempts to include relevant baryonic physics have found mechanisms that may be effective in erasing a NFW-type dark matter cusp, or even preventing one from ever forming. At least four such mechanisms have been proposed: (1) rapid removal ("blowout") of a large quantity of central gas due to a starburst causing the dark matter to expand [e.g., 89], and energy and angular momentum transfer to the central dark matter through the action of (2) bars [90], (3) gas motion (e.g. [91]), and (4) infalling clumps via dynamical friction [92-94]. Proposal (1) is supported by recent cosmological simulations of formation of small spiral galaxies (F. Governato et al., in preparation). Recent high-resolution simulations [95] do not favor (2). But recent work has suggested (3) that supernova-driven gas motions could smooth out dark matter cusps in very small forming galaxies as a consequence of resonant heating of dark matter in the fluctuating potential that results from the bulk gas motions [96], and thus explain observations suggesting dark matter cores in dwarf spheroidal (dSph) galaxies such as the Fornax and Ursa Minor satellites of the Milky Way. These authors suggest that the same mechanism can explain other puzzling features of dSph galaxies, such as the stellar population gradients, the low decay rate for globular cluster orbits, and the low central stellar density. They also argue that once the dark matter cusp is smoothed out by baryonic effects in protogalaxies, subsequent merging will not re-create a cusp even in larger galaxies [cf. 97]. AGN-driven bulk gas motion has also been shown to be a possible explanation for dark matter and stellar cores in massive stellar spheroids [98].

Recent work also suggests (4) that dynamical friction could explain the origin of dark matter cores in dwarf spheroidal galaxies [99,100] and in low-mass disk galaxies [101,102]. The latter papers compare ΛCDM pure dark matter (PDM) and dark matter + baryons (BDM) simulations starting from the same initial conditions consistent with WMAP3 cosmological parameters. The hydrodynamic BDM simulation includes star formation and feedback. At high redshifts $z > 7$, the PDM and BDM density profiles are very similar. Adiabatic contraction [103-108] subsequently causes the BDM halo to become more cuspy than the PDM one, but then dynamical friction causes infalling baryon+DM clumps to transfer energy and angular momentum to the dark matter. The resulting DM radial profile is essentially pseudo-isothermal with a flat core – see the low-$z$ curves in Fig. 2: in the inner ~2 kpc, $\rho(R)$ becomes flat (left panel). The behavior of the dark matter velocity dispersion $\sigma_{DM}$ in the PDM vs. BDM models mirrors that of the density. The NFW cusp in the PDM simulation forms early and is characterized by a "temperature inversion": $\sigma_{DM}(R)$ rising to R ~10 kpc.

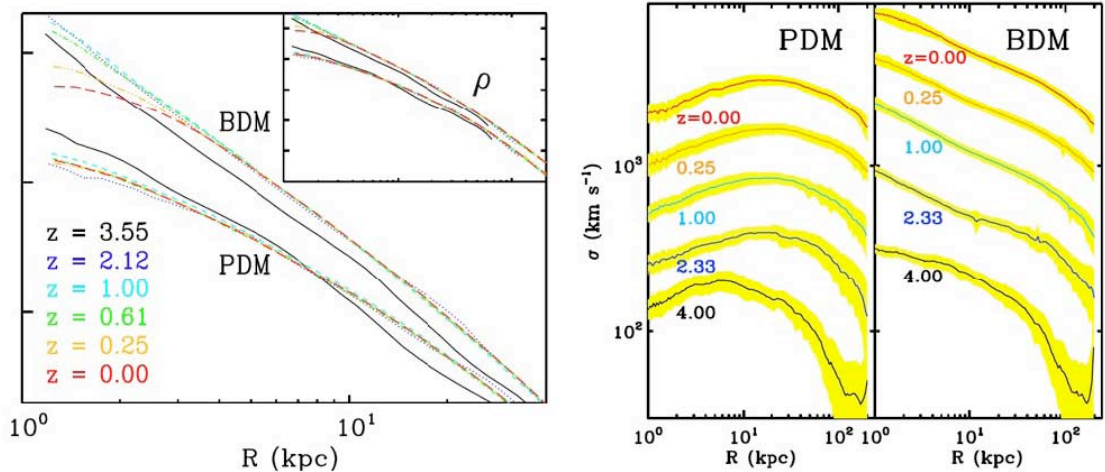

**FIGURE 2. (left)** Redshift evolution of DM density profiles $\rho(R)$ in PDM and BDM models: z = 3.55 (solid), 2.12 (dotted), 1.0 (dashed), 0.61 (dot-dashed), 025 (dot-dash-dotted) and 0 (long dashed). The PDM and BDM curves are displaced vertically for clarity. The inner 40 kpc of halos are shown. The vertical coordinate units are logarithmic and arbitrary. For the PDM model, the density is well fitted by the NFW profile over a large range in z, and $r_s$~28 kpc at z = 0. For the BDM model, the NFW fit is worse and $R_{iso}$~15 kpc at the end. The insert provides $\rho$ within 200 kpc for comparison. **(right)** Redshift evolution of DM velocity dispersions in PDM and BDM models. Except for the lowest ones, the curves are displaced vertically up for clarity. The second curves from the bottom are displaced by a factor of 2, the third — by a factor of $2^2$, the fourth — by a factor of $2^3$, and the last ones— by a factor of $2^4$. The colored width represents a 1σ dispersion around the mean. The inner 200 kpc of halos are shown. The vertical coordinate units are logarithmic. (From [101].)

But in the BDM simulation there is no temperature inversion, and indeed $\sigma_{DM}(R)^2 \sim R^{-\beta}$ with β increasing until about $z \sim 0.6$ and decreasing sharply thereafter; this is apparently caused by dynamical friction heating the central DM, causing it to stream outward. The number of subhalos in this inner region of the BDM simulation is about twice that of the PDM simulation, which could be relevant for explaining the anomalous flux ratios in radio quads (discussed in the previous section). The central density distribution in the BDM simulation may be what is needed to explain strong lensing statistics [109]. These very intriguing simulation results need to be confirmed and extended by higher resolution simulations of many more galaxies.

Observations indicated that dark matter halos may also be too concentrated farther from their centers [110] compared to ΛCDM predictions. Halos hosting low surface brightness galaxies may have higher spin and lower concentration than average [111,80], which would improve agreement between ΛCDM predictions and observations. As we have just discussed, it remains unclear how much adiabatic contraction [103-106] occurs as the baryons cool and condense toward the center, since there are potentially offsetting effects from gas motions [96] and dynamical friction [101]. Recent analyses comparing spiral galaxy data to theory conclude that there is little room for adiabatic contraction [112,113], and that a bit of halo expansion may better fit the data [113]. Early ΛCDM simulations with high values $\sigma_8 \sim 1$ of the linear mass fluctuation amplitude in spheres of 8 $h^{-1}$ Mpc (a measure of the amplitude of the power spectrum of density fluctuations) predicted high concentrations [114], which are lower with lower values of $\sigma_8$ [115]. The cosmological parameters from WMAP5 and large scale structure observations [2], in particular $\sigma_8 = 0.82$, lead to concentrations that match galaxy observations better [116], and they may also match observed cluster concentrations [117,118]. (For recent work on concentration vs. redshift, see [119].)

## ANGULAR MOMENTUM ISSUES

The growth of the mass of dark matter halos and its relation to the structure of the halos has been studied based on structural merger trees [111], and the angular momentum of dark matter halos is now understood to arise largely from the orbital angular momentum of merging progenitor halos [120,121]. But it is now clear that the dark matter and baryonic matter in disk galaxies have very different angular momentum distributions [122,123]. Although until recently simulations were not able to account for the formation and structure of disk galaxies, simulations with higher resolution and improved treatment of stellar feedback from supernovae are starting to produce disk galaxies that resemble those that nature produces [124,125]. It remains to be understood how the gas that forms stars acquires the needed angular momentum. High-resolution hydrodynamical simulations also appear to produce thick, clumpy rotating disk galaxies at redshifts $z > 2$ [126], as observed [127,128]. Possibly important is the fairly recent realization that, rather than being heated to the halo virial temperature as in the standard treatment used in semi-analytic models [1,129], a significant amount of gas enters halos cold and in clouds or streams [130-133] in halos less massive than $\sim 10^{12}$ M$_\odot$, or even in more massive halos at $z > 2$.

Once thin stellar disks form, they are in danger of being thickened by mergers. One expects major mergers to be more common for larger mass galaxies because the increasing inefficiency of star formation in higher mass halos limits the total stellar masses of galaxies [134]. Studies of mergers in simulations show that for Milky Way mass galaxies, the largest contribution in mass comes from mergers with a mass ratio of ~1:10 [135]. Thin disks are significantly thickened by such mergers [136], although if the merging galaxies are gas rich, a relatively thin disk can re-form [137-139]. That the majority of large mergers onto halos less massive than ~$10^{12}$ $M_\odot$ are gas rich while the gas fraction decreases for more massive halos >$10^{12.5}$ $M_\odot$ [140] could help to explain the increasing fraction of large stellar spheroids in larger mass halos [141]. In the absence of good statistics on the disk thickness of galaxies and the relative abundance of bulgeless disks as a function of galaxy mass, the Sérsic index is a useful proxy. In the Milky Way mass range ($V_{rot} \approx 220$ km s$^{-1}$, $M_{star} \sim 10^{11}$ $M_\odot$) less than 0.1% of blue galaxies are bulgeless, while for M33-mass galaxies ($V_{rot} \approx 120$ km s$^{-1}$, $M_{star} \sim 10^{10}$ $M_\odot$) bulgeless galaxies are more common, with 45% of blue galaxies having Sérsic index $n < 1.5$. Thus the challenge for ΛCDM is to produce enough M33-type galaxies [142].

## SMALL SCALE ISSUES: SUMMARY

**Satellites**: The discovery of many faint Local Group dwarf galaxies appears to be consistent with ΛCDM predictions. Reionization, lensing, satellites, and Lyman-alpha forest data imply that if the dark matter is WDM, it must be tepid at most – i.e., not too warm.

**Cusps:** Recent high-resolution observations of nearby low-mass disk galaxies provide strong evidence that the central dark matter often has a nearly constant density core, not the NFW-type $\rho(r) \propto r^{-1}$ cusp. But the target is changing (which no doubt infuriates some observers), as high-resolution ΛCDM simulations including baryons appear to be producing dwarf spheroidal and low-mass spiral galaxies consistent with these observations. Better observations and simulations are needed.

**Angular Momentum:** ΛCDM simulations are increasingly able to form realistic spiral galaxies, as resolution improves and feedback is modeled more physically. However, accounting for the statistics on thin disks and bulgeless galaxies as a function of galaxy mass will be a challenge for continually improving simulations and semi-analytic models.

## ACKNOWLEDGMENTS

I thank NASA and NSF for grants that supported research relevant to this topic. I also thank my current and former students and other colleagues, including the participants in the 2009 Caltech workshop "Shedding Light on the Nature of Dark Matter" (http://www.kiss.caltech.edu/mini-study/darkmatter/index.html) supported by the W. M. Keck Institute for Space Studies (KISS), for many helpful discussions.